\documentclass[runningheads,a4paper]{llncs}

\usepackage{amssymb}
\usepackage{graphicx}

\usepackage{url,comment}
\urldef{\mailsa}\path|{hector.zenil,|
\urldef{\mailsb}\path|narsis.kiani,|
\urldef{\mailsc}\path|jesper.tegner}@ki.se|    
\newcommand{\keywords}[1]{\par\addvspace\baselineskip
\noindent\keywordname\enspace\ignorespaces#1}

\begin{document}

\mainmatter  

\title{Numerical Investigation of Graph Spectra and Information Interpretability of Eigenvalues}

\titlerunning{Numerical Investigation of Graph Spectra}

%
%
\author{Hector Zenil, Narsis A. Kiani and Jesper Tegn\'er
%
}
\authorrunning{Zenil et al.}

\institute{Unit of Computational Medicine, Department of Medicine,\\
Centre for Molecular Medicine, Karolinska Institute\\Stockholm, Sweden.\\
\mailsa\\
\mailsb\\
\mailsc\\
\url{http://www.compmed.se}}

%
%

\toctitle{Numerical Investigation of Graph Spectra and Information Interpretability of Eigenvalues}
\tocauthor{Numerical Investigation of Graph Spectra and Information Interpretability of Eigenvalues}
\maketitle



\begin{abstract}
We undertake an extensive numerical investigation of the graph spectra of thousands regular graphs, a set of random Erd{\"o}s-R\'enyi graphs, the two most popular types of complex networks and an evolving genetic network by using novel conceptual and experimental tools. Our objective in so doing is to contribute to an understanding of the meaning of the Eigenvalues of a graph relative to its topological and information-theoretic properties. We introduce a technique for identifying the most informative Eigenvalues of evolving networks by comparing graph spectra behavior to their algorithmic complexity. We suggest that extending techniques can be used to further investigate the behavior of evolving biological networks. In the extended version of this paper we apply these techniques to seven  tissue specific regulatory networks as static example and  network  of a na{\"i}ve pluripotent immune cell in the process of differentiating towards a Th17 cell as evolving example, finding the most and least informative Eigenvalues at every stage.
\end{abstract}

\section{Background}

The analysis of large networks raises in many of research fields, the ubiquity of large networks  makes the analysis of the common properties of these networks important.In the most simplistic way can be seen or analyzed as a collection of vertices and edges but there are a  very different way of representing the graph, using the eigenvalues and eigenvectors of matrices associated with the graph (Graph Spectra) rather than the vertices and edges themselves.In this study  a graph or network $G$ defined by pairs $(V(G),E(G))$,where $V(G)$ is a set of  vertices (or nodes) and $E(G)$ represent edges(links). Let $A$ be an $n \times n$ real matrix. An eigenvector of $A$ is a vector such that $Ax = \lambda x$ for some real or complex number $\lambda$. $\lambda$ is called the Eigenvalue of $A$ belonging to Eigenvector $v$. The set of graph Eigenvalues of the adjacency matrix is called the spectrum of the graph. Spectral analysis is a widely used  for a range of problems. In general, assigning meaning to Eigenvalues is very difficult. They are very context sensitive (i.e. relative to the graph type) and they are cryptic in the sense that they store many properties of a graph in a single number that does not lend itself to being easily used to reconstruct the properties it encodes. However, they are known to encode algebraic and topological information relating to a graph in various ways. In this paper we contribute toward the investigation of the interpretability of Eigenvalues, specifically with a general method to determine the type and the amount of information about a network that each Eigenvalue carries. We analyse growing networks ranging from  complete graphs to complex random network and demonstrate the distinct behaviour of the eigenvalue spectra of different topology class. We will show the unique spectral properties of  the major random graph models, \textit{Erd{\"o}s-R\'enyi}~\cite{erdos,gilbert}, \textit{small-world}~\cite{watts}  and \textit{scale free}~\cite{albert2002statistical}.

\section{Methodology}

All graphs in this paper are undirected, so that the matrices are symmetrical and the Eigenvalues are real. They also have no loops, so the matrices have a zero diagonal and hence a zero trace, so that the Eigenvalues add up to zero. We are interested in investigating the behavior of $Spec(G)$ relative to the Kolmogorov complexity $K(G)$. Formally, the Kolmogorov complexity of a string $s$ is $K(s)=\min\{|p| : U(p)=s\}$. That is, the length (in bits) of the shortest program $p$ that when running on a universal Turing machine $U$ outputs $s$ upon halting. A universal Turing machine $U$ is an abstraction of a general-purpose computer that can be programmed to reproduce any computable object, such as a string or a network (e.g. the elements of an adjacency matrix). By the \emph{Invariance theorem}~\cite{li}, $K_U$ only depends on $U$ up to a constant, so as is conventional, the $U$ subscript can be dropped. Formally, $\exists \gamma$ such that $|K_U(s) - K_{U\prime}(s) | < \gamma$ where $\gamma$ is a constant independent of $U$ and $U\prime$. Due to its great power, $K$ comes with a technical inconvenience (called \textit{semi-computability}) and it has been proven that no effective algorithm exists which takes a string $s$ as input and produces the exact integer $K(s)$ as output~\cite{kolmo,chaitin}. Despite the inconvenience $K$ can be effectively approximated by using, for example, compression algorithms. Kolmogorov complexity can alternatively be understood in terms of uncompressibility. If an object, such as a biological network, is highly compressible, then $K$ is small and the object is said to be non-random. However, if the object is uncompressible then it is considered algorithmically random.

\subsubsection{Algorithmic probability}
\label{bdm}

There is another seminal concept in the theory of algorithmic information, namely the concept of \textit{algorithmic probability}~\cite{solomonoff,levin} and its related \textit{Universal distribution}, also called Levin's \textit{probability semi-measure}~\cite{levin}. The algorithmic probability of a string $s$ provides the probability that a valid random program $p$ written in bits uniformly distributed produces the string $s$ when run on a universal (prefix-free\footnote{The group of valid programs forms a prefix-free set (no element is a prefix of any other, a property necessary to keep $0 < m(s) < 1$.) For details see~\cite{cover,calude}.}) Turing machine $U$. In equation form this can be rendered as $m(s) = \sum_{p:U(p) = s} 1/2^{|p|}$. That is, the sum over all the programs $p$ for which $U$ outputs $s$ and halts. The algorithmic Coding Theorem~\cite{levin} establishes the connection between $m(s)$ and $K(s)$ as $|-\log_2 m(s) - K(s)| < \mathcal{O}(1) \textnormal{ (Eq. 1)}$, where $\mathcal{O}(1)$ is an additive value independent of $s$. The Coding Theorem implies that~\cite{cover,calude} one can estimate the Kolmogorov complexity of a string from its frequency by rewriting Eq.~(1) as $K_m(s)=-\log_2 m(s) + \mathcal{O}(1) \textnormal{ (Eq. 2)}$.

\begin{figure}[htbp!]
\centering
\includegraphics[width=5.7cm]{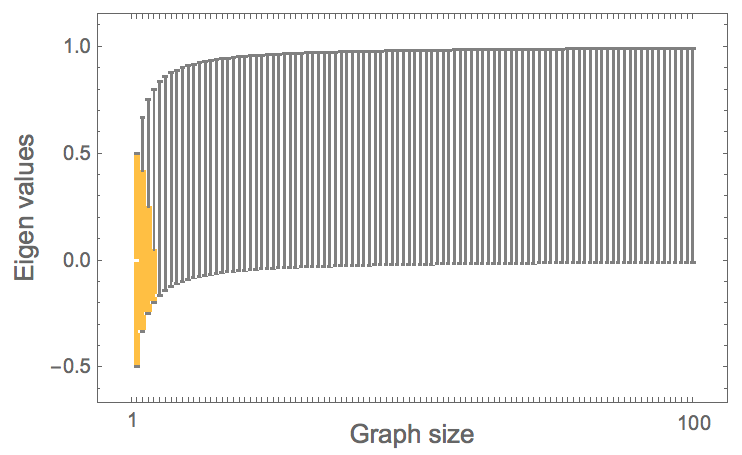} \hspace{.5cm} \includegraphics[width=5.7cm]{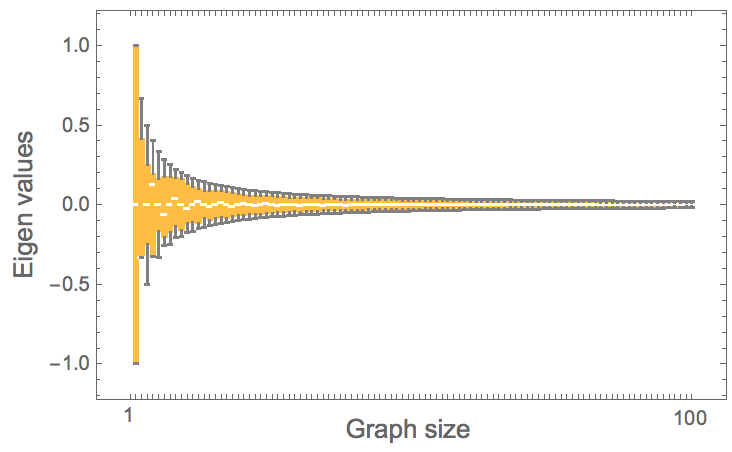}\\

\bigskip
\includegraphics[width=5.7cm]{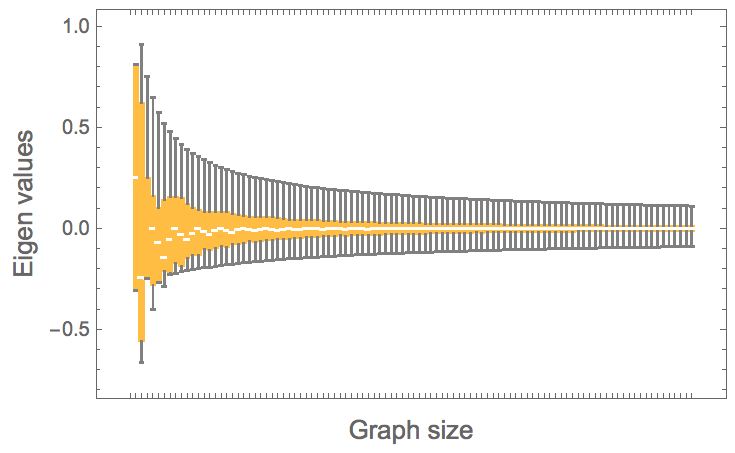} \hspace{.5cm} \includegraphics[width=5.7cm]{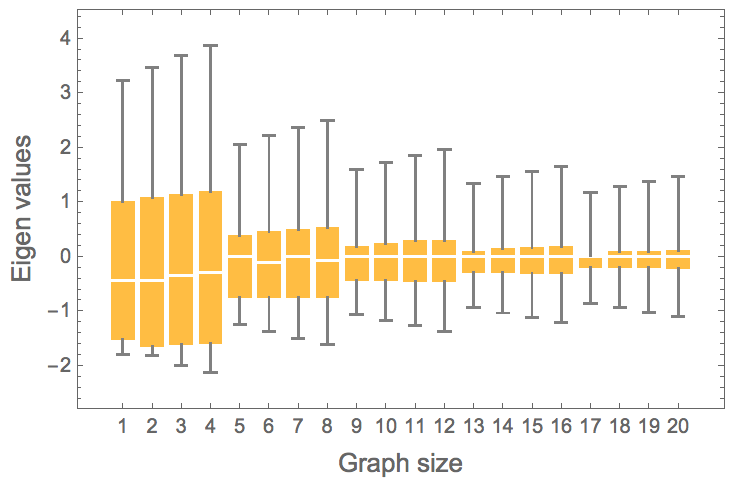}\\

\bigskip
\includegraphics[width=5.7cm]{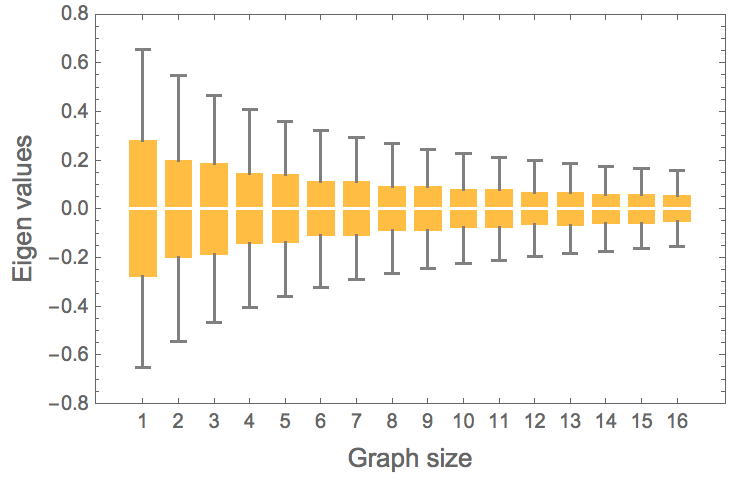} \hspace{.5cm} \includegraphics[width=5.7cm]{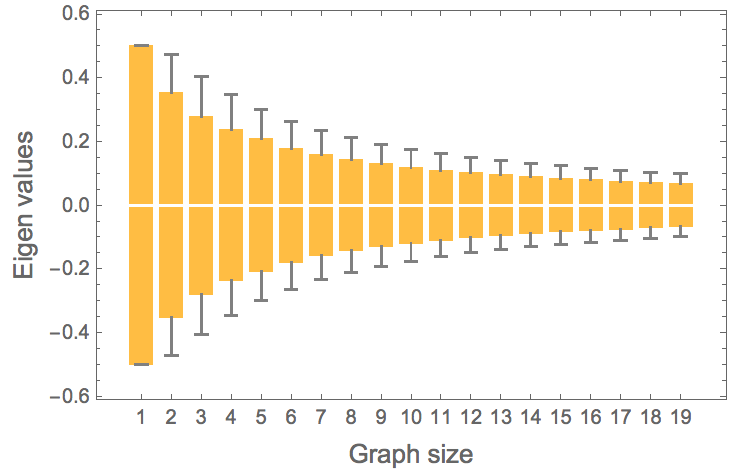}
\caption{Box distributions of Eigenvalues for growing regular graphs normalized by edge count. We call these ``spectra signatures". Top left: Spectra signature of a growing complete graph showing that the Eigenvalues normalized by edge count do not carry any extra information than may be found in a simple Kolmogorov complexity graph. Top right: Spectra signature of a growing cycle graph showing a wider range of different Eigenvalues centered around $x=0$. The next spectra signatures have an increasing number of different Eigenvalues but remain relatively simple given the regular structure of the graphs they represent. The diversity of Eigenvalues can be captured by classical Shannon entropy, but the non-trivial structure can only be captured by algorithmic complexity. Middle left: Spectra signature of a growing wheel graph. Middle right: Spectra signature of a growing fan graph. Bottom left: Spectra signature of a growing lattice graph. Bottom right: Spectra signature of a growing path graph. Obvious similarities between similar graphs can be recognized: cycles and wheels have similar patterns, grids and paths share some similarities too. However, star and fan graphs have spectra that show a greater degree of disparity than the spectra of the others.}
\label{spectrasignatures}
\end{figure}

\subsubsection{Kolmogorov complexity of Unlabeled graphs}

As shown in~\cite{zenilgraph}, estimations of Kolmogorov complexity may be arrived at by means of the algorithmic Coding theorem, using a 2-dimensional lattice as tape for a 2-dimensional deterministic universal Turing machine. Hence $m(G)$ is the probability that a random computer program acting on a 2-dimensional grid prints out the adjacency matrix of $G$. Essentially it uses the fact that the more frequently an adjacency matrix is produced, the lower its Kolmogorov complexity  and vice versa. We call this the \textit{Block Decomposition Method} (BDM) as it requires the partition of the adjacency matrix of a graph into smaller matrices using which we can numerically calculate its algorithmic probability by running a large set of small 2-dimensional deterministic Turing machines, and thence, by applying the algorithmic Coding theorem, its Kolmogorov complexity. Then the overall complexity of the original adjacency matrix is the sum of the complexity of its parts, albeit with a logarithmic penalization for repetitions, given that $n$ repetitions of the same object only adds $\log n$ to its overall complexity. Formally, the Kolmogorov complexity of a labeled graph $G$ by means of $BDM$ is defined as $K_{BDM} (G,d) = \sum_{(r_u,n_u)\in A(G)_{d\times d}} \log_2(n_u)+K_m(r_u)$, where $K_m(r_u)$ is the approximation of the Kolmogorov complexity of the subarrays $r_u$ by using the algorithmic Coding theorem (Eq.~(2)), and $A(G)_{d\times d}$ represents the set with elements $(r_u,n_u)$ obtained when decomposing the adjacency matrix of $G$ into non-overlapping squares of size $d$ by $d$. In each $(r_u,n_u)$ pair, $r_u$ is one such square and $n_u$ its multiplicity (number of occurrences). From now on $K_{BDM} (g,d=4)$ will be denoted only by $K(G)$ but it should be taken as an approximation to $K(G)$ unless otherwise stated (e.g. when taking the theoretical true $K(G)$ value). More details of these measures and their application are given in~\cite{zenilgraph}. The Kolmogorov complexity of a graph $G$ is thus given by:

$$
K^\prime(G)=\min\{K(A(G_L)) | G_L \in L(G)\}
$$

\noindent where $L(G)$ is the group of all possible labelings of $G$ and $G_L$ a particular labeling. In fact $K(G)$ provides a choice for graph canonization, taking the adjacency matrix of $G$ with lowest Kolmogorov complexity. Unfortunately, there is almost certainly no simple-to-calculate universal graph invariant, whether based on the graph spectrum or any other parameters of a graph. In~\cite{newzenilgraph}, however, we proved that the calculation of the complexity of any labeled graph is a good approximation to its unlabeled version. 


\section{Results}

\subsection{Most informative Eigenvalues}

\begin{figure}[htbp!]
\centering
\includegraphics[width=2.7cm]{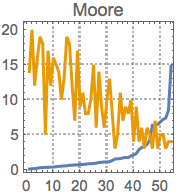}
\includegraphics[width=2.7cm]{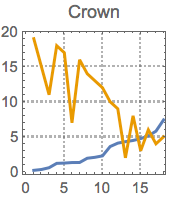}
\includegraphics[width=2.7cm]{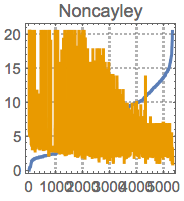}
\includegraphics[width=2.7cm]{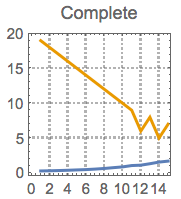}
\includegraphics[width=2.7cm]{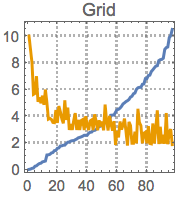}
\includegraphics[width=2.7cm]{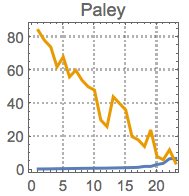}
\includegraphics[width=2.7cm]{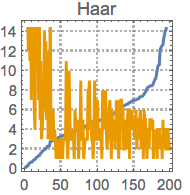}
\includegraphics[width=2.7cm]{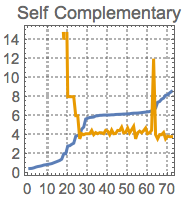}
\caption{Correlation plots of graph complexity vs largest Eigenvalues. On the $X$-axis are graphs (blue/darker curve) sorted by their algorithmic complexity (from lower to higher information content) normalized by graph edge count. On the $Y$-axis are the largest Eigenvalues for each graph (yellow/lighter curve). Both complexity and Eigenvalues are normalized by graph edge count as we are interested in structural information contained in both measures beyond information about the graph size.}
\label{plot1}
\end{figure}

\begin{figure}[htbp!]
\centering
\includegraphics[width=2.7cm]{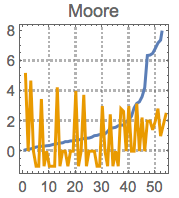}
\includegraphics[width=2.7cm]{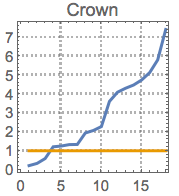}
\includegraphics[width=2.7cm]{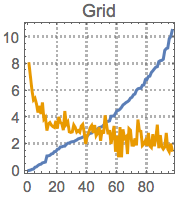}
\includegraphics[width=2.8cm]{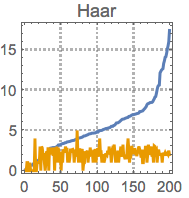}
\caption{Correlation plots of graph complexity vs second largest Eigenvalues. The second largest Eigenvalue displays a larger variety of correlations among graph classes, and depicted here is a case where it is found that the second value does not carry any information about Crown graphs. For Moore graphs the positive correlation is weak, and for Haar graphs it is null but noisy, unlike for Crowns. Specific statistics are given in Fig.~\ref{correlation} quantifying the correlations across all graph classes.}
\label{plot2}
\end{figure}

\begin{figure}[htbp!]
\centering
\includegraphics[width=2.75cm]{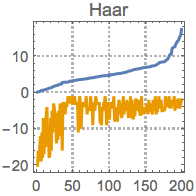}
\includegraphics[width=2.6cm]{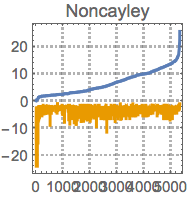}
\includegraphics[width=2.8cm]{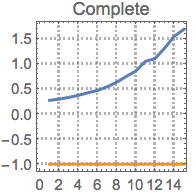}
\includegraphics[width=2.7cm]{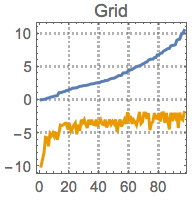}
\includegraphics[width=2.7cm]{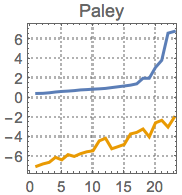}
\includegraphics[width=2.9cm]{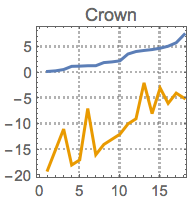}
\includegraphics[width=2.8cm]{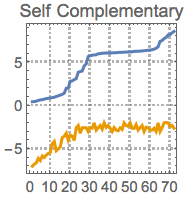}
\caption{Correlation plots of graph complexity vs smallest Eigenvalues. Smallest Eigenvalues tend to be positively correlated to graph information content. Depicted here is again a known example of a non-informative Eigenvalue for complete graphs, which is nonetheless informative in the sense that deleting the effect of size from its information content retrieves almost no information, hence all Eigenvalues and the complexity of the graph are basically flat (notice $Y$-axis scale). In another example, unlike the second largest Eigenvalue, it can be seen that the smallest Eigenvalue does carry information about Crown graphs.}
\label{plot3}
\end{figure}

\begin{figure}[htbp!]
\centering
\includegraphics[width=6cm]{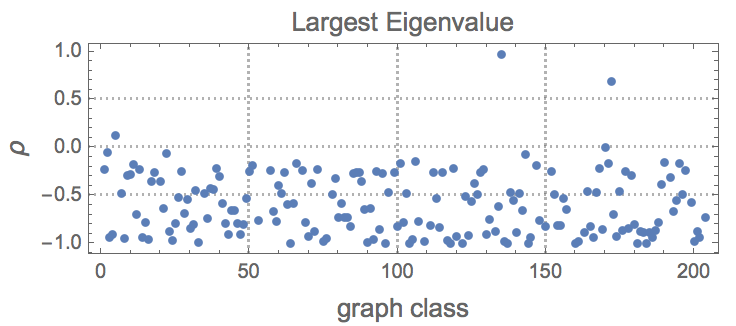} \vspace{.5cm} \includegraphics[width=6cm]{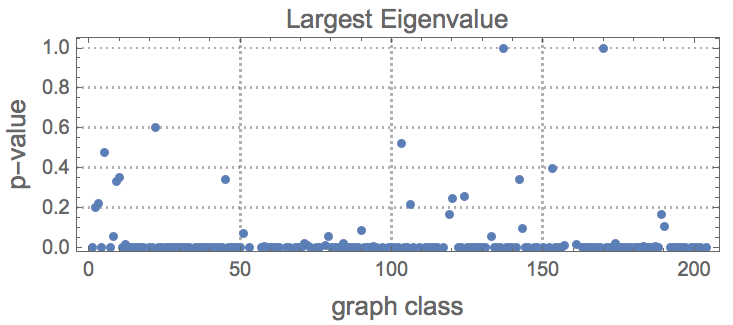}\\

\includegraphics[width=6cm]{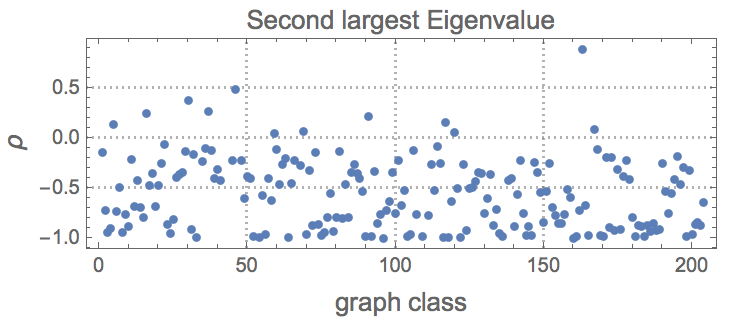} \vspace{.5cm} \includegraphics[width=6cm]{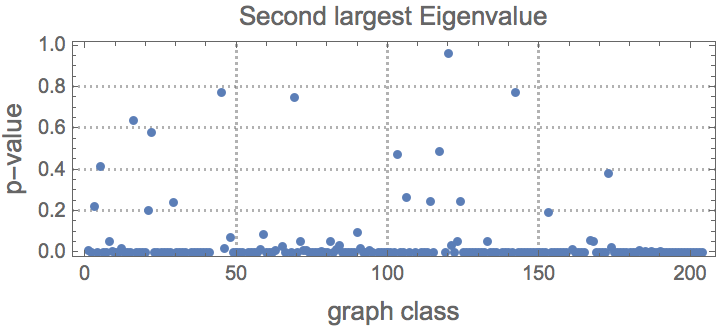}\\

\includegraphics[width=6cm]{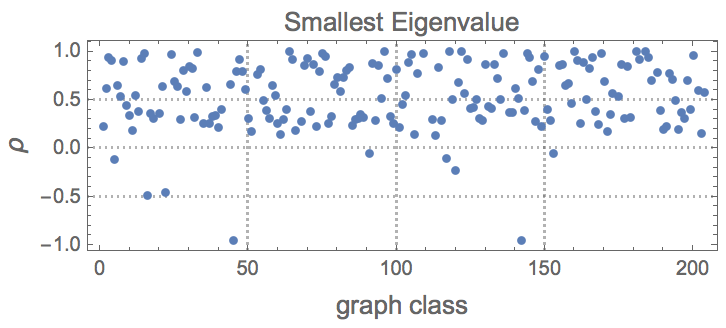} \vspace{.5cm} \includegraphics[width=6cm]{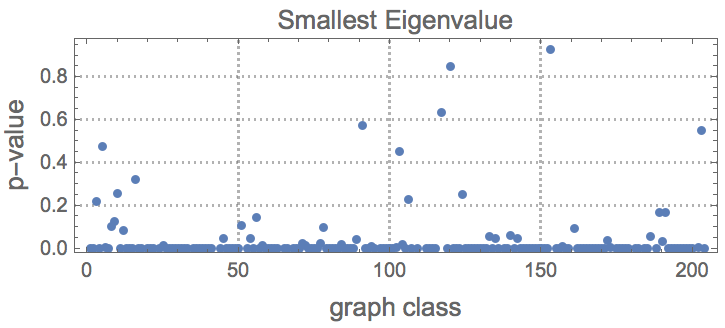}\\

\caption{Statistics ($\rho$) and p-value plots between graph complexity and largest, second largest and smallest Eigenvalues of 204 different graph classes including 4913 graphs. Clearly the graph class complexity correlates in different ways to different Eigenvalues but in most cases this correlation is strong and there is a clear tendency of the largest Eigenvalue to be negatively correlated to information content, then a quick transition at the second largest and finally a clear positive correlation with the smallest.}
\label{correlation}
\end{figure}

It is clear that Eigenvalues carry different information and therefore can be of differential informative value. For example, take a complete graph of size $n$. To reconstruct it from its graph spectra it is enough to look at its largest Eigenvalue $\lambda_1$, simply because it indicates the size of the complete graph and therefore contains all the information about it-- assuming that we know it is a complete graph. If we did not know it to be a complete graph then we would need to take into account the rest of the $n$ Eigenvalues, but none of them on its own would suffice. That is only if a graph with $\lambda_1 \neq 0$ and $\lambda_i = -1$ with $i=2$ to $n$ uniquely determines a complete graph.

In Figs.~\ref{plot1}, \ref{plot2}, \ref{plot3} and~\ref{correlation}, a sample of 4913 graphs distributed in 204 classes dividing (with possible repetition) the networks into bins of shared topological or algebraic properties, such as being a Moore, Haar, Cayley, tree or acyclic graph, display various (mostly significant) degrees of negative and positive correlation with one or more Eigenvalues. The number of graphs come from the graphs available in the \textit{Mathematica} v.10 software built-in repository function GraphData[]. The most commonly found case was a negative correlation between largest Eigenvalues and graph information content. However, positive correlation and non-trivial differences between next largest and smallest Eigenvalues were found and their behavior is highly graph-topology dependent. This suggests that while the largest Eigenvalue encodes important structural information of the graph, all Eigenvalues may carry some information, with some being more or less informative than others. The complete graph is a trivial example of no correlation, where it is clear that the Eigenvalue is not providing any structural information about the graph other than its size, which is erased when normalized by edge count as it is in these plots, hence discounting by any edge count contribution. The degree and type of correlation can be found in Fig.~\ref{correlation}, quantified by a typical Pearson correlation test.

If the Eigenvalue behavior of a graph $G$ is flat, then its information-content is low or null, except perhaps because of the multiplicity of the value and the total number of occurrences of the same value, trivially indicating, for example, the size of the network, given that the number of Eigenvalues is equal to the number of vertices of $G$. This also means that Eigenvalues with flat behavior are less informative, a fact which enables clear discrimination between interesting and uninteresting Eigenvalues, beyond a simple consideration of numerical value (numerical values can be different and still not carry any information about a graph).

\subsection{Graph spectra behavior of evolving networks}

\begin{figure}[htbp!]
\centering
\includegraphics[width=12cm]{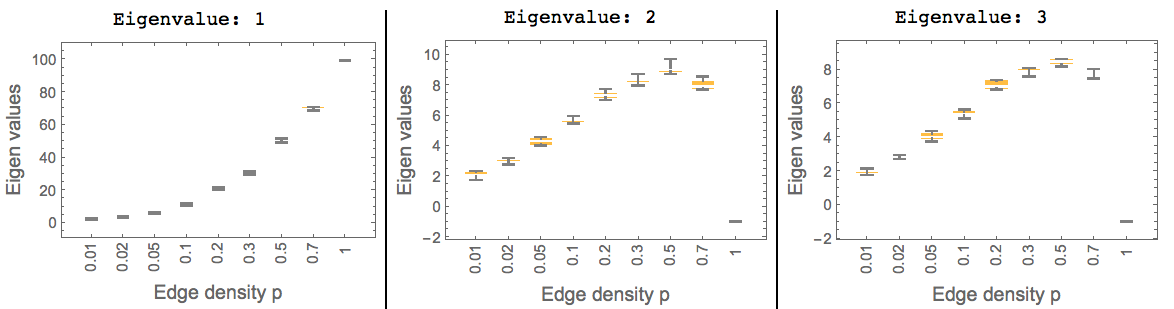}
\caption{Eigenvalues behavior. The largest Eigenvalue in a random E-R graph of size 100 vertices for edge density from 0 to 1 ($X$-axis) is the only one behaving differently from the rest. Some properties of the largest Eigenvalue are known, such as being an indicator of number of bifurcations, so the greater the edge count the greater its value. However, the next Eigenvalues all manifest a common behavior, reaching a maximum and describing a concave curve.}
\label{randomspectradensity}
\end{figure}


\begin{figure}[htbp!]
\centering
\includegraphics[width=12cm]{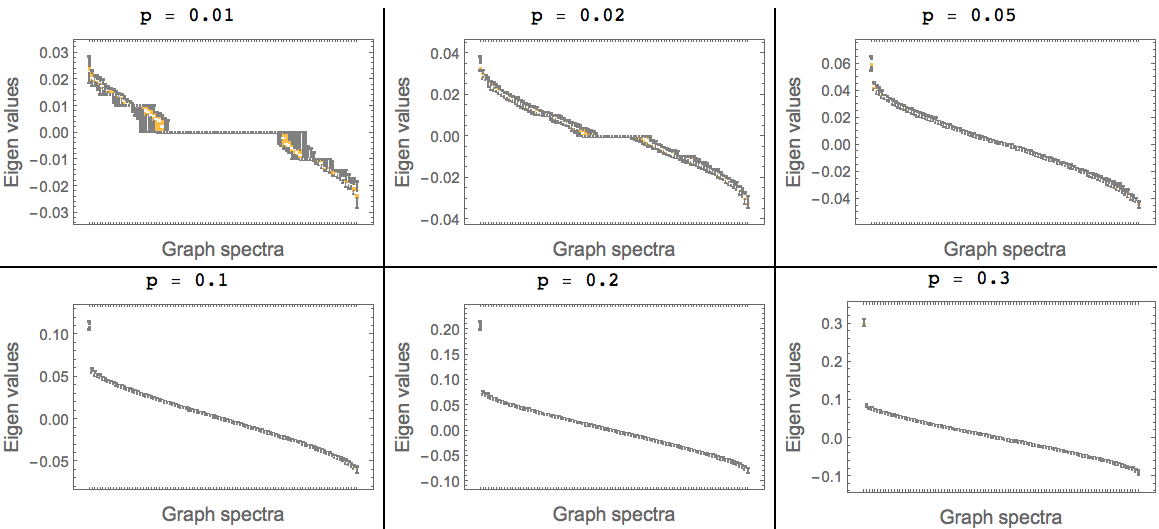}
\caption{Spectra signature of a random E-R graph of size 100 for edge density 0 to 1. Clearly for edge density 1, the random graph spectra are simply those of a complete graph.}
\label{randomspectrasignature}
\end{figure}

\begin{figure}[htbp!]
\centering
\includegraphics[width=12cm]{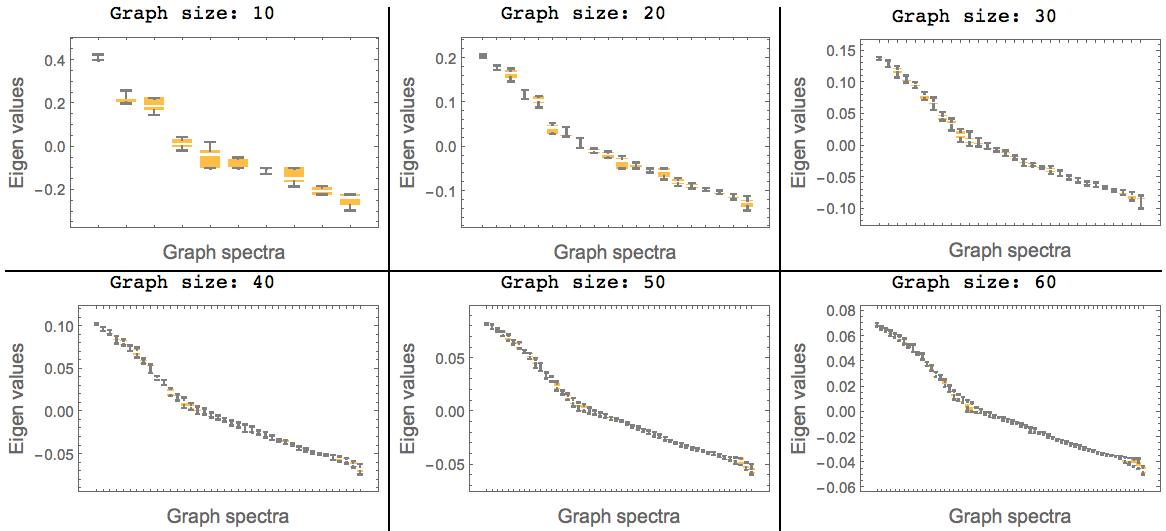}
\caption{Spectra signature of a Watts-Strogatz growing into a 100-node network with rewiring probability 0.05.}
\label{wsspectrasize}
\end{figure}

\begin{figure}[htbp!]
\centering
\includegraphics[width=12cm]{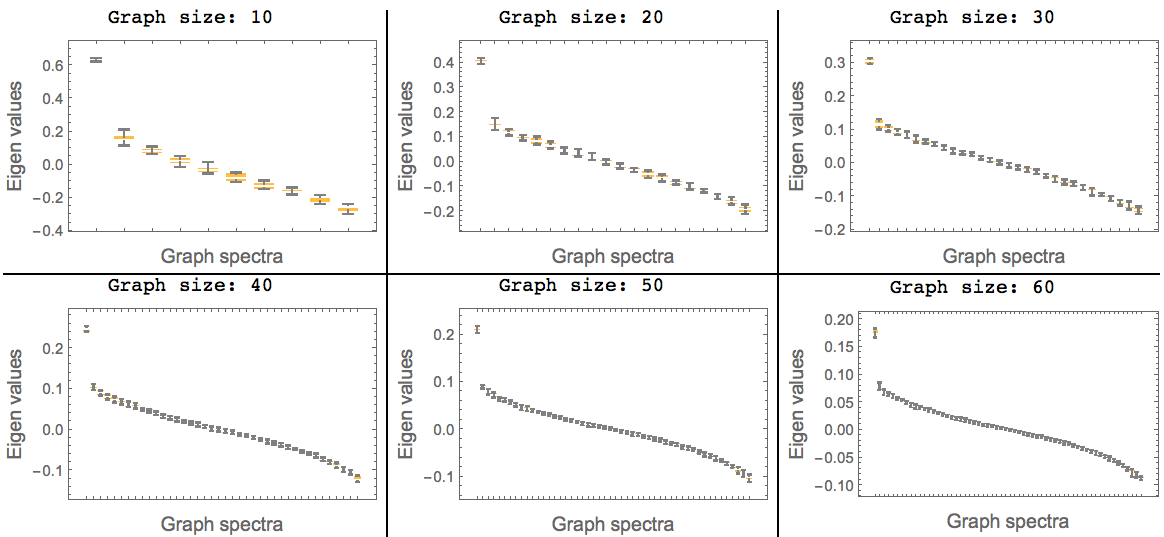}
\caption{Spectra signature of a growing Barab\'asi-Albert network reaching a size of 100 nodes where a new vertex with 4 edges is added at each step.}
\label{baspectrasize2}
\end{figure}

\subsection{Spectra signatures}

We compared the \textit{spectra signature} of an evolving graph to the Box plots of the Eigenvalues of the graph over time. Fig.~\ref{spectrasignatures}, for example, shows the asymptotic behavior of each Eigenvalue for well known regular graphs and how the plots characterize them with various regular patterns, including cyclic behavior for a cycle graph. They also show how the accumulation of Eigenvalues is distributed differently for different graphs, with their rate of growth depending on the graph type. A complete graph $G$ of size $n=|V(G)|$, for example, has graph spectra $(n -1)^1, (-1)^{n-1}$ with its values corresponding to the plot in Fig.~\ref{spectrasignatures}(left).
When the number of different Eigenvalues is small (i.e. their multiplicity is too high) and they converge soon to a fixed normalized Eigenvalue, this is an indication that the Eigenvalue carries no information or is exhausted after a few evolving steps (i.e. no more information can be extracted, or the graph can be characterized after a few evolving steps) (see spectra signatures in Fig.~\ref{spectrasignatures}). We undertook a novel numerical investigation of the Eigenvalues of growing graphs for different classes, shedding light on both known and possibly unexplored properties of Eigenvalues for some specific graph types. To this end we calculated what we defined as spectra signatures of random and complex networks prior to a deeper investigation concerning the information content of synthetic graphs and biological networks.

\section{Conclusions}

We have introduced a concept of spectra signatures based upon numerical calculations of growing networks with different group-theoretic and topological properties for the study of evolving network behavior. We have then moved toward the information content of these networks via estimating their Kolmogorov complexity by means of entropy, lossless compression and algorithmic probability (BDM).

We have introduced an analysis based on correlation comparisons of each Eigenvalue against the information content of a graph to reveal the most informative Eigenvalue for different graph classes. We found that the largest Eigenvalues are negatively correlated to graph complexity even after edge count normalization, while the smallest Eigenvalues are in general not correlated or positively correlated, with only a couple of cases of negative correlation. While most research has focused on a few of the largest Eigenvalues of a graph spectrum, we have shown that in actual fact the smallest Eigenvalues carry a high information content as often as the largest.
The techniques introduced here can be extended to Laplacian matrices, but Laplacian matrices carry only redundant information about the degree of the vertices because the original graph can be reconstructed from the adjacency matrix alone. Thus the effect of $Spec(G)$ on the Laplacian or simple spectra of $G$ with respect to $K(G)$ is negligible. For Kolmogorov complexity, we have $|K(A_L(G)) - K(A(G))|<c$, where $A_L(G)$ is the Laplacian matrix of $G$, $A(G)$ is the simple adjacency matrix of $G$ and $c$ is the algorithm implementing the Laplacian calculation $L = D(G) - A(G)$, where $D(G)$ is the diagonal degree matrix of $G$.
We believe this is a novel approach to extracting meaning from and thus contributing to the solution of the problem of the interpretability of graph spectra, a fundamental step toward applications of graph spectra theory in network biology, especially in the context of evolving networks--given that some biological models are represented as Ordinary Differential Equations for which this approach, when applied to the Jacobian matrices of the ODEs, is thoroughly relevant. As introduced here, this approach promises to be able to reveal specifics about the behavior of a biological network over time through the study of Eigenvalues in relation to their information-content.

One future research direction is the investigation of behavioral differences in Eigenvalues of networks representing disease as compared to those of healthy networks, both as profiling techniques and as a tool for understanding the direction in which a healthy network may over time progress towards a disease state.

\end{document}